\documentclass[12pt,letterpaper]{article}
\usepackage[top=1.4in,bottom=1.4in,left=0.80in,right=0.80in]{geometry}
\usepackage[hidelinks]{hyperref} 
\usepackage{amsmath}
\usepackage{amssymb}
\usepackage{imakeidx}
\usepackage{authblk}
\usepackage{textcomp}		
\usepackage{graphicx,caption}
\graphicspath{\Images}
\usepackage{caption}
\usepackage{subcaption}
\usepackage[super,comma,compress]{natbib} 
\usepackage[utf8]{inputenc}
\usepackage{lscape}
\usepackage{pgfgantt}
\usepackage{geometry}
\usepackage{tikz}
\usepackage{pgfplots}
\usetikzlibrary{shadings}
\usepackage{comment}
\usepackage{color}
\pgfplotsset{compat=1.18}
\usepackage{mathtools}
\usepackage{ulem}

\newcommand{\mrr}{\color{black}}

\date {}
\begin{document}
\title{
Life and Death of a Thin Liquid Film
}


\author{Muhammad Rizwanur Rahman${^1}$, Li Shen${^1}$, James P. Ewen ${^1}$, D. M. Heyes${^1}$, Daniele Dini${^1}$, and E. R. Smith${^2}$}
\affil{${^1}$Department of Mechanical Engineering, Imperial College London, South Kensington Campus, London SW7 2AZ, United Kingdom \newline ${^2}$Department of Mechanical and Aerospace Engineering, Brunel University London, Uxbridge UB8 3PH, United Kingdom}

\maketitle

\begin{abstract}

Thin films, bubbles and membranes are central to numerous natural and engineering processes, i.e., in thin-film solar cells, coatings, biosensors, electrowetting displays, foams, and emulsions.
Yet, the characterization and an adequate understanding of their rupture is limited by the scarcity of atomic detail.
We present here the complete life-cycle of freely suspended films using non-equilibrium molecular dynamics simulations of a simple atomic fluid free of surfactants and surface impurities, thus isolating the fundamental rupture mechanisms.
Counter to the conventional notion that rupture occurs randomly, we discovered a short-term ‘memory’ by rewinding in time from a rupture event, extracting deterministic behaviors from apparent stochasticity. 
A comprehensive investigation of the key rupture-stages including both unrestrained and frustrated propagation is made - characterization of the latter leads to a first-order correction to the classical film-retraction theory.
Furthermore, the highly resolved time window reveals that the different modes of the morphological development, typically characterized as heterogeneous nucleation and spinodal decomposition, continuously evolve seamlessly with time from one into the other.

\end{abstract}

\pagebreak 
\section*{Introduction}

Isaac Asimov's \textit{Foundation} series introduced the concept of \textit{psychohistory} - where historical trends enable statistical forecasts of the overall trajectories of the civilization while not predicting individual or transient events. 
In terms of rupture and nucleation dynamics, this parallels the notion of being microscopically stochastic yet macroscopically deterministic - an important distinction as we work through a complete description of the multi-scale rupture process of a thin liquid film.

The stability and rupture of thin films holds vast significance in a wide spectrum of applications \citep{duran2019instability,craster2009dynamics}, ranging from natural processes like gravity currents, lava flows and snow avalanches \citep{ancey2007plasticity,goldstein2014instability} to biological transport processes in lungs and cell membranes, and in disease transmission and forensic analyses \citep{grotberg2001respiratory, villermaux2020fragmentation,tammaro2021flowering}, to more common place engineering applications in miniaturized micro-electronic, micro-fluidic and biomedical devices \citep{griesser2016thin, piegari2018optical}, complex coatings, distillations and insulation \citep{stone2004engineering, eijkel2005nanofluidics} - to name a few.
Consequentially, the literature concerning the stability and rupture of thin liquid films is voluminous - encompassing theoretical \citep{scheludko1962certaines,kashchiev1980nucleation,derjaguin1981theory,vaynblat2001rupture,saulnier2002dewetting}, experimental \citep{evers1996rupture,nikolova1999rupture,casteletto2003stability}, numerical \citep{shen2013stability} and molecular dynamics \citep{gamba1992molecular,bresme2004computer,jang2006structures,tarazona2012newton} investigations.

The rupture process, which becomes inevitable upon a film's sufficient thinning, requires an activation energy of $\mathcal{O}(\gamma h^2)$ \citep{de1958foam}, where $h$ and $\gamma$ denotes, respectively, the thickness and the surface tension of a film. Although this renders this mechanism improbable for thick films \citep{vrij1966possible}, 
the probability of rupture 
increases as thickness decreases, and becomes most significant for films thinner than {\it ca.}
$10\,\mbox{nm}$. 
The spontaneity in the rupture process is mostly triggered by the growth of fluctuations and corrugations of the surface due to thermal motion that critically deteriorates the structural integrity of the film, leading to rupture. 
Thermal undulation assumes increasing importance for films with nanoscopic thickness, as highlighted by the molecular dynamics investigation of \citet{zhang2019molecular}, and the undulations reduce the critical wavelength causing rupture of otherwise (classically) stable thin films.

In addition to the continuous distortion of the liquid-vapor interface due to thermal fluctuations \citep{langevin2020rupture}, a slight (local) gradient in surface tension immobilizes the interface \citep{yaminsky2010stability} inducing fluid movement, though it tends to be overlooked in the majority of theoretical models.
The considerations of uniform thickness and a stationary interface, as in the study of \citet{derjaguin1981theory}, result in a rupture thickness within the range of a few hundred nanometers. 
In contrast, \citeauthor{anderson2010foam} \citep{anderson2010foam} conducted a linear stability analysis of thinning films, showing that rupture actually occurs when the film thins down to a few tens of nanometers. 
Yet, the linear theories cannot follow the film evolution  
until rupture since their validity ceases as the disturbances in the film grows sufficiently.
Inclusion of the non-linear terms showed that any local thinning of the film amplifies the influence of the long-range force, while simultaneously attenuating the effects of surface tension \citep{williams1982nonlinear} . 
The importance of relaxing the linear approximations in the film stability and rupture studies is further realized by the fact that the non-linear theory yields significantly shorter time of rupture from those obtained through the linear models, as much as an order of magnitude \citep{williams1982nonlinear}. 
Despite the extensive efforts in investigating film stability and rupture, a comprehensive understanding, however, remains elusive owing to the intricate and interrelated physical processes governing rupture across scales. Particularly the unfolding of rupture from its origin, and the fundamental difference between the rupture modes remain unclear - partly due to the experimental challenge at the scale where it initiates, and partly due to
the fundamental complexity which predominantly resides in the nature of the interconnected spatio-temporal dynamics of the liquid-vapor surface - which constitutes the focus of interest of this study.

The central topic of the current investigation concerns the complete life-span of thin (black) films, and their rupture mechanism. By unveiling the fundamental aspects and the behavior of thin films at the smallest of scales and mapping the mechanisms responsible for the evolution of films across space-time coordinates, we provide, for the first time, a unified view of the evolution of thin-film dynamics. This is achieved by tracing back and explaining the links between inception and fragmentation phases, and by identifying the structural similarities emerging at different scales.
The subsequent sections of this manuscript are structured as follows: 
we begin by identifying the shared characteristics among various modes of rupture, and unify them.
Following this, we demonstrate a time-sequencing behavior of the film in the moments leading up to its rupture.
This deterministic aspect within the typically stochastic rupture process underscores the general principle that the interplay between stochastic micro-scale events and deterministic macro-scale outcomes is essentially a common thread across diverse fields of study. 
Finally, we provide detailed illustrations of the nucleation event,
the late stage frustrated propagation leading to coalescence. The latter gives fresh insights on factors responsible for slower propagation and, hence, delayed coalescence.

\section*{Results}
\begin{figure}[!ht]
    \centering
    \includegraphics[width=0.98\linewidth]{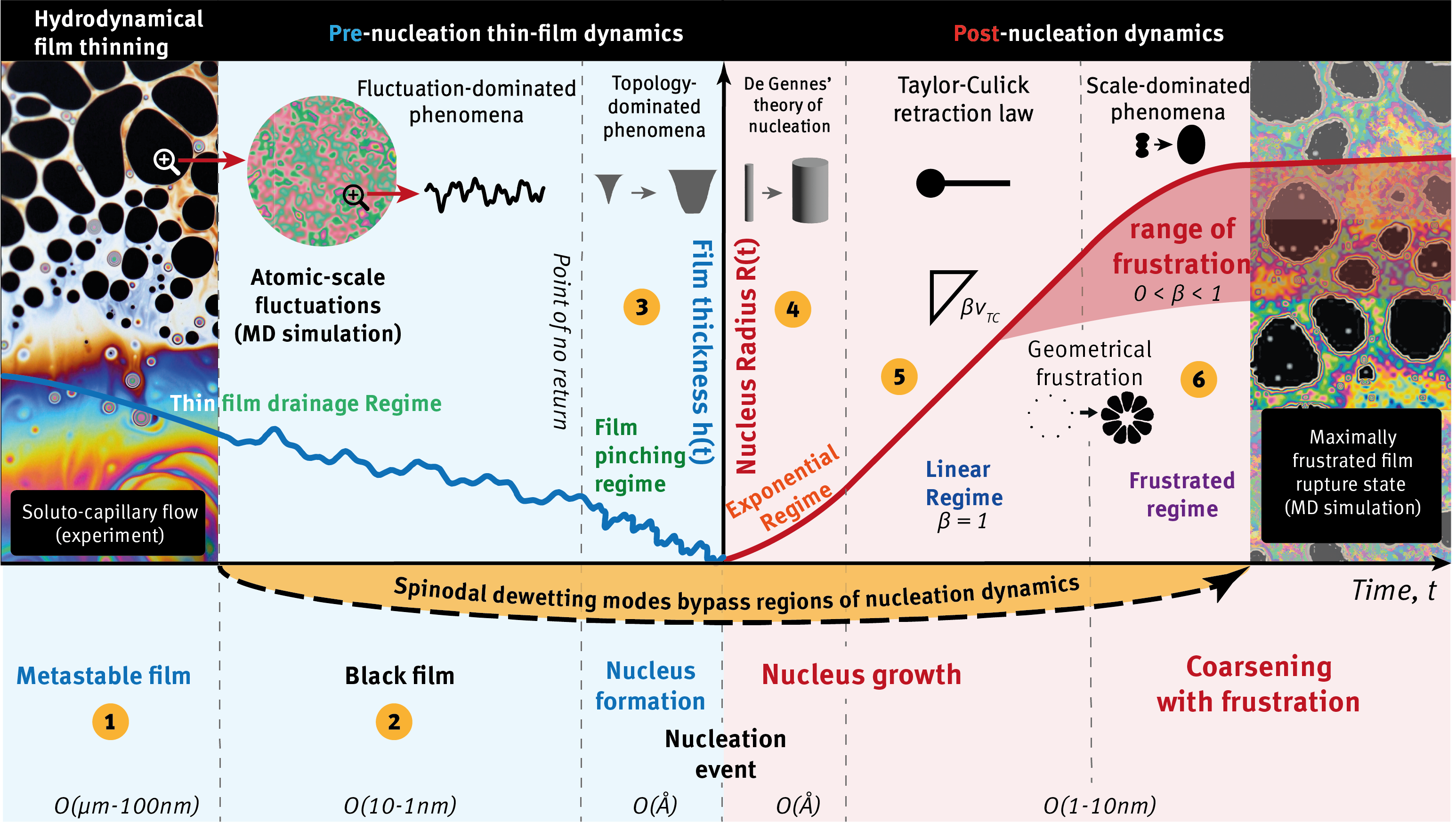}
    \caption{Portrait of the life-span of a thin-film. The left half of the schematic, regions 1-3, shows thinning in stages before nucleation, the y-axis represents film thickness (not to scale). The far-left (region-1) snapshot shows (experimental) hydrodynamic thinning in a micron-thickness metastable films which leads to the formation of nanoscopic black films. Remaining parts of the schematic summarizes molecular dynamics observations which are  inaccessible by experimental investigation: (region-2) fluctuation dominated thinning of black films, (region-3) as rupture becomes imminent, thinning is accelerated and leads to film pinching and eventually to nucleation when $h=0$. The just-nucleated site now starts expanding. Subsequent growth of the ruptured site is presented in the right half of the figure (shaded in faint red, regions 4-6)  with the y-axis now representing the radius of the rupture site. (region-4) Initially growth is exponential which soon attains (region-5) linear growth. As more sites rupture across the film, their presence and growths frustrate the growth of any individual site as in (region-6) which later follows coalescence and coarsening - the far right snapshot from MD simulations shows a typical state of the (black) film at this stage with expanding holes in it. The spinodal rupture is fast-forwarded from 1 to 6, even at the MD scale, (almost) bypassing the intermediate stages.}
    \label{fig:schematic}
\end{figure}

The rupture process is long preceded by the drainage and subsequent thinning of the metastable film. 
Thinning is driven by capillary suction, which is followed by the development of an instability leading to rupture. Such an instability is caused by the heterogeneity in film thickness, or due to the gradients in surface concentration~\citep{saulnier2014study,de2001some}. 
Fig.~\ref{fig:schematic} summarizes the full life-time of a film: starting from hydrodynamic thinning at micron thickness (region 1 in Fig.~\ref{fig:schematic}) and subsequent thinning at nanoscale (region 2 in Fig.~\ref{fig:schematic}), followed by a piercing stage when surface fluctuations grow more rapidly at some sites (or, locations) compared to others leading to nucleation (region 3 in Fig.~\ref{fig:schematic}). 
Depending on the initial film thickness, however, the rupture process may follow one of the several modes of nucleation; and although a stochastic process, a finite window of memory is observed - these we discuss in the next two sections.  
Once the film is punctured, a short term exponential growth regime emerges (region 4 in Fig.~\ref{fig:schematic}) when the nucleus attains a circular shape and eventually enters the linear growth regime (region 5 in Fig.~\ref{fig:schematic}). 
This linear growth, referred to as the Taylor-Culick retraction, has been extensively investigated in the literature, both numerically and by experiment; and has been studied for freely suspended Lennard-Jones (LJ) films in an earlier study by the authors, also through molecular dynamics simulations as here, see \citet{rahman2023non} and references therein. 
Depending on the modes of nucleation and the film thickness, multiple rupture sites may nucleate and continue to grow. 
This growth is later affected by the presence and growth of neighboring nuclei, and a frustrated growth regime emerges (region 6 in Fig.~\ref{fig:schematic}) leading to delayed coalescence and coarsening. These stages will be discussed separately in sections to follow.

\subsection*{Modes of Rupture}\label{sec:Modes}
\begin{figure}[!th]
    \centering
    \includegraphics[width=0.85\linewidth]{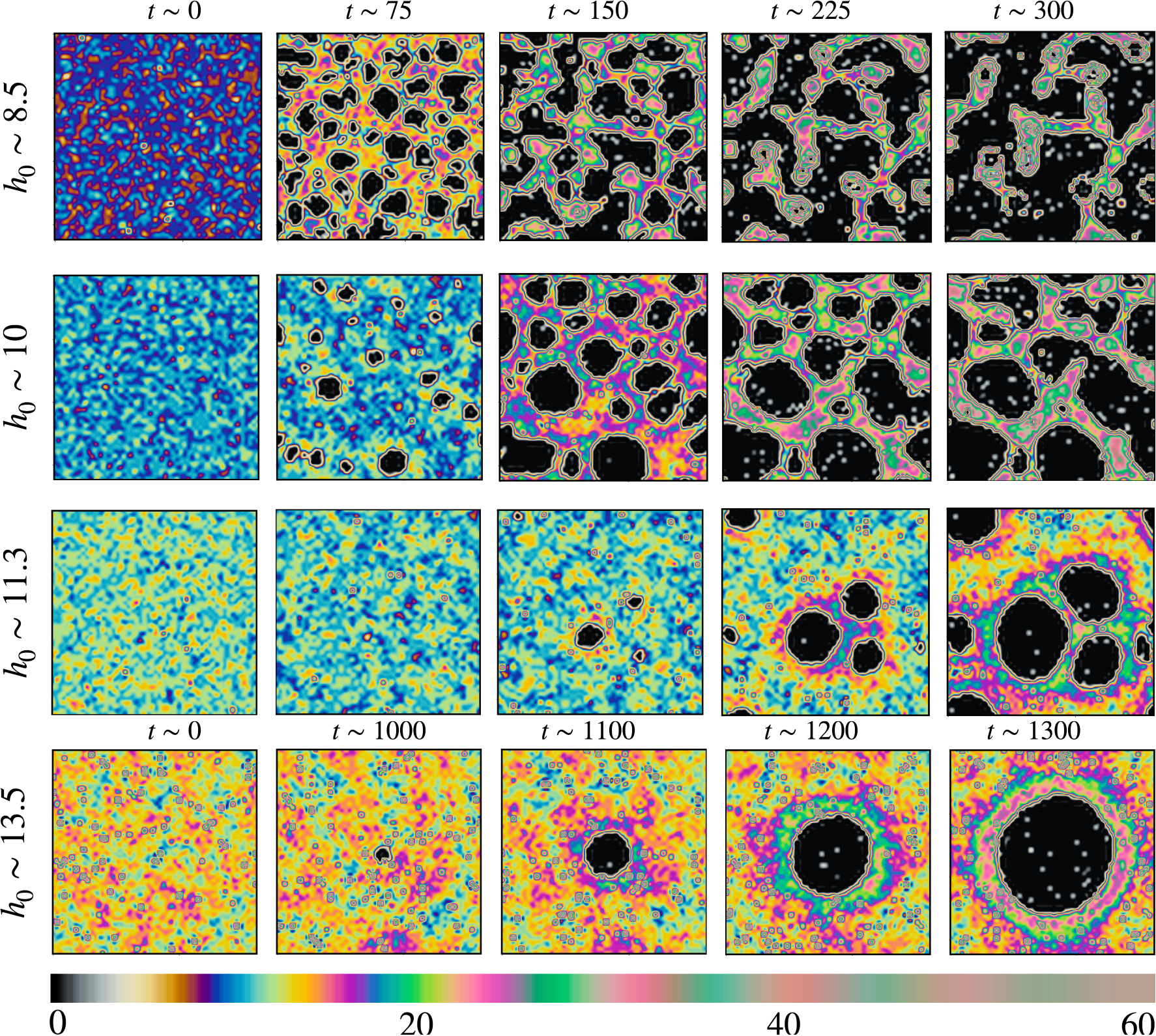}
    \caption{Spontaneous rupture of films of different initial thicknesses, $h_0 = 8.5, 10, 11.3$ and $13.5$. Color maps represent the local thickness. The presence of surface modulations are seen in the unruptured states of the films.}
    \label{fig:both_modes} 
\end{figure} 
\noindent {\textbf{Spinodal Dewetting vs. Heterogeneous Nucleation:}} 
In a freely suspended film (and also those on a substrate), \textit{spinodal dewetting}~\cite{vrij1966possible,xie1998spinodal,higgins2000anisotropic,thiele2001dewetting,Becker2003} refers to its solely thermodynamic and rapidly spontaneous disintegration, in the absence of an energy barrier, as the film moves towards a lower energy state, with irregular but self-similar and uniformly distributed random patterns in a region with a gradient in composition. 
As a film grows thicker, the increasing relevance of the bulk fluid forms an energy barrier which the film needs to overcome in order to nucleate. In regions with no sharp compositional gradients, a slower kinetic or thermodynamic \textit{homogeneous nucleation} process results in a uniformly distributed nuclei-pattern. 
If a film has surface impurities or defects, then the energy barrier is (locally) lowered and the uniformly distributed nuclei patterns become sporadic in space and in time, which is commonly referred to as the \textit{heterogeneous nucleation}. 
Due to the lack of an energy barrier, spinodal dewetting is the fastest mode of disintegration, with heterogeneous nucleation in second place due to the presence of an energy barrier albeit weakened by local defects, and finally the full energy barrier results in the  slowest mode of disintegration, i.e., homogeneous nucleation which is typical of films with thickness much above the atomistic length scale.

The present results give evidence of the existence of  spinodal dewetting and the heterogeneous nucleation modes in freely suspended thin films. 
Fig.\,\ref{fig:both_modes} shows snapshots of the film thickness at different times 
for varying initial thickness, $h_0$. The thinnest film ($h_0 \sim 8.5$ Lennard Jones units, top row) considered in this study displays uniformly distributed nuclei with a narrow size distribution and bi-continuous surface patterns $-$ typical characteristics of spinodal dewetting. 
The transition from a `vapor-in-liquid' to a `liquid-in-vapor' state is evident, and the irregular shapes and sizes of the nuclei arise not only from early coalescence but also from the interactions between neighboring nuclei. 
As the film thickness increases (from top to bottom panels), the uniformity of the location of rupture sites diminishes, the nuclei become more circular.
A few nuclei become sufficiently larger than others - a typical characteristic assigned to heterogeneous nucleation~\citep{reiter1994dewetting, stange1997nucleation,thiele1998dewetting}. 
These observations agree with the experimental investigation with polystyrene films on silicon substrates by \citeauthor{xie1998spinodal}\,(\citeyear{xie1998spinodal}) who proposed a crossover thickness from spinodal to heterogeneous nucleation.

The coexistence of both of the modes discussed above was also observed for evaporating structural protein films \citep{thiele1998dewetting}, for liquid metals \citep{bischof1996dewetting, herminghaus1998spinodal}, and for polymer films \citep{xie1998spinodal, jacobs1998thin, seemann2001dewetting}. 
However, these studies were based on liquid films on solid substrates, and to the best of our knowledge, the rupture of freely suspended single component films has not been systematically investigated until now.
In the current study, the observations of rupture for various thicknesses agree qualitatively with the commonly observed patterns found for spinodal and heterogeneous nucleation; but most importantly, these are found to initiate spontaneously without the presence of surface impurities or external perturbations. 
It is clear from the present free film model that
the so-called `heterogeneous nucleation' observed in this study must arise from highly localized thermal fluctuations, rather than any substrate defect as in \citet{thiele1998dewetting} and elsewhere.

Upon nucleation within a comparatively thicker film, an individual nucleus undergoes expansion at the Taylor-Culick speed, creating a rim rich in particles encircling it \citep{rahman2023non}. This rim serves as a deterrent to the growth and coalescence of approaching neighbor nuclei, and thereby delays coalescence (which we discuss later in further detail).
When an individual nucleus finds it energetically more favorable to amalgamate than to continue to grow, coalescence takes place.
In contrast, the dynamics differ for spinodal dewetting where nucleation occurs `almost' simultaneously at multiple sites, creating a uniform distribution across the film. 
Consequently, the nuclei face spatial constraints, hindering their expansion. 
Additionally, spinodal dewetting typically takes place in relatively thinner films with higher nuclei growth rates, facilitating easy expulsion of the liquid bridge trapped between neighboring nuclei resulting in an earlier coalescence stage. 

Notably, these observations highlight that the transition between these modes can occur solely based on the thickness of the film and on the spread of the surface instability, indicating the critical role played by the film thickness in determining the dominant dewetting mechanism. 
This, therefore, expands the remit of the present results potentially to other substrate-free situations. In addition, it can be inferred that any apparent dissimilarity in the rupture patterns of two films of different $h_0$ must arise from the spatio-temporal distribution of the nucleation events. If these differences in the distribution of nucleation events are removed, one obtains self-similar evolution across scales - from growth to fragmentation (see {\mrr{\textit{SI: section 1}}} for further details).

\subsection*{Memory of Rupture}\label{sec:Memory}
\begin{figure}[!h]
    \centering
    \includegraphics[width=.85\linewidth]{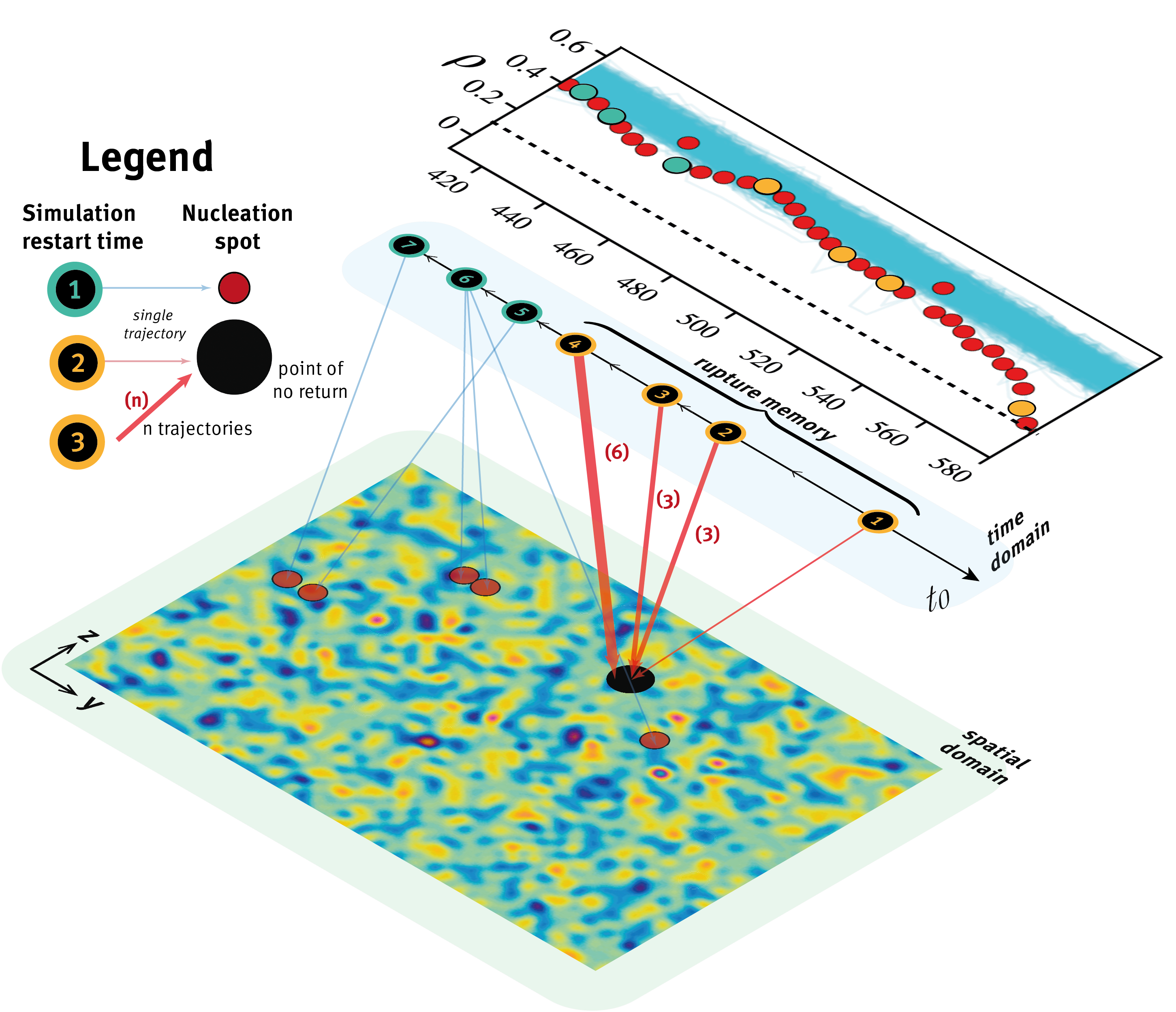}
    \caption{Memory of rupture: (top) temporal variation of the thickness-averaged local density. The circles corresponds to density of a location where nucleation occurs at $t\sim580$, whereas, the cyan curves shows local density at other locations on the film which do not nucleate within the time plotted here. 
    Several independent simulation were restarted at time $t=418,449,481,504,520, 571$ - these time instances are marked by the circles on the time axis, and the corresponding density of the ruptured site at those time instants are marked by blue and orange circles.  
    On the snapshot of the film at the bottom, the position of nucleation for every case  is depicted either by a black circle if the location of nucleation remains the same as of the original case, or red if the film nucleates at a different location. 
    Similarly, red lines connect nuclei when films nucleate at the same spot as the mother film, and blue lines connect otherwise. $(n)$ shows number of independent simulations (also represented by thickness of the connecting lines), omitted if $n=1$.}
    \label{fig:film-memory}
\end{figure}

The probabilistic nature of film rupture process prohibits any precise prediction of the time and location of a nucleation-event, especially for a defect free surface. 
Perhaps surprisingly, our study uncovers a short-term window before the nucleation event where it becomes deterministic - even with a perturbation applied to the trajectories. And this sets the scene for our discussion on the state of MDMS, i.e. macroscopically deterministic, microscopically stochastic events. 
For the nucleation of thin liquid films, this is the state where the first hole will nucleate  in a specific bounded region $\Omega$ of space, almost surely, regardless of the number of times the scenario is repeated whereby each individual repetition produces a slight alternation of the trajectory of film evolution. In Fig.\,\ref{fig:film-memory}, $\Omega$ is schematically represented by the circular nucleation spot. 
Moreover there exists a brief time-window, $\Sigma$, shown in Fig.\,\ref{fig:film-memory} as the rupture memory region, within which we can rewind the film state and still achieve nucleation in $\Omega$. 
Together, the region defined by $\Omega \times \Sigma$ forms the spatio-temporal boundary for the MDMS characteristics of a film approaching its first nucleation.

To explore this temporal duration of impending rupture events, we systematically preserved the system's state at regular intervals and subsequently resumed the simulation from those instances onward. Repetition of these computer experiments highlights the inherent weak extent of `memory' in the film-rupture process. 
This means that, under identical initial conditions, the time and location of rupture exhibit variability in independent simulations, except for an exceptionally narrow time interval during which the film displays deterministic behavior.
Fig.\,\ref{fig:film-memory} elucidates this dynamics. 
The top plot illustrates density fluctuations over time, with filled red symbols representing the local density at the center of a rupture site that nucleates at $t \sim 580$. 
The cyan curves portray local density at all other film positions, showcasing temporal fluctuations through the curve bandwidth. 
Throughout this extensive time span, despite the density of the rupture site closely following the lower boundary of the density spectrum, no abnormal deviations relative to other density profiles occur.
Hence, assigning a critical lower density threshold would be inaccurate. This is supported by the observation of the density profiles at certain regions that do indeed experience significant density reduction (see the cyan density profile that approaches $\rho < 0.2$ at $t \sim 520$ and still heals), they do not culminate in rupture/nucleation events. 

The time instances denoted by the orange or blue circles on the density plot (and, also on the time-axis encircling the black circles) mark the moments when independent simulations were re-started. These instances are reproduced, alongside the time axis, in the lower panel, where the $y-z$ coordinates of the film are depicted, pinpointing nucleation sites represented by filled circles. Connection lines link each restart time to its corresponding nucleation location. 
The larger black circle designates the nucleation site of the original trajectory ($t_n \sim 580$, as displayed in the top panel's). Notably, in a simulation recommenced from an earlier state, i.e., $t_0 \sim 520$, the film still nucleated at the same spot. This consistent location remained unaltered until $t_0 < 481$.
The case (re)started from $t_0 \sim 481$ is particularly important as one can identify a rise in the local density which implies that a relatively thicker area on the film can be more prone to rupture than a thinner area.
To corroborate this observation, multiple independent simulations were conducted around $t_0 \sim 481$. Even when introducing minor perturbations to the initial trajectory, the film consistently underwent nucleation at the identical site.
Restarting the simulation from a state earlier than $\sim480$ introduced a distinct behavior, where the film undergoes nucleation at random and diverse locations and times. 
These stochastic nucleation sites are depicted as faint red circles in the figure, linked by blue lines to corresponding restart times. 
Among these latter cases, when restarting from the state at $t_0 \sim 449$, one of the cases coincidentally nucleated at the same location as the original simulation, which we regard as an incidental occurrence.
A similar rewind is conducted for another case for which $t_n \sim 450$, see {\mrr\textit{SI: section 2}}, the film is observed to nucleate at a different location and at a different time, but exhibits similar MDMS characteristics with a different spatio-temporal boundary - suggesting reduced, if not complete absence of, stochasticity\,\citep{chatzigiannakis2020breakup} within the memory-window.

As illustrated above, local density alone cannot define the likelihood of rupture. And neither does the local energy nor the local temperature - 
we observed no noticeable difference in those parameters immediately before the rupture event, and elect not to present here for the sake of brevity.
This leaves one possible explanation for the `spatio-temporal' memory of film rupture.
We can interpret the transition from the probabilistic to the deterministic stages of nucleus (or, hole) formation/evolution
in terms of Stillinger's inherent structure
theory of dynamical evolution in liquids, which has been widely applied to many aspects of liquid dynamical phenomena at the molecular level \citep{stillinger1982hidden, nusse1994basins, bagchi2013water, ball2014hidden}.
In this theory it is proposed there is a so-called `basin of attraction' formed from all those equilibrium molecular configurations - which when quenched, instantaneously collapse into the {\textit{same}} underlying or `inherent' structure, some of
which we have seen are precursors to nucleus formation. 
Therefore, the trajectories (re)started from different points in time, despite their microscopic randomness, are perhaps still attached to the same inherent structure and results in the same outcome.
If one goes too far back in time, slight differences in the trajectory associated with the restart may cause the system to evolve in a way where it becomes associated with or `captured' by another inherent structure which may not lead to a nucleus (unlike the situation before the rewind). 
In the present context, certain inherent structures can be viewed as acting as `gateways' to hole formation.

\subsection*{Formation, Frustration and Coalescence}\label{sec:Coalescence}

\begin{figure}[!t]
    \centering
    \includegraphics[width=0.95\linewidth]{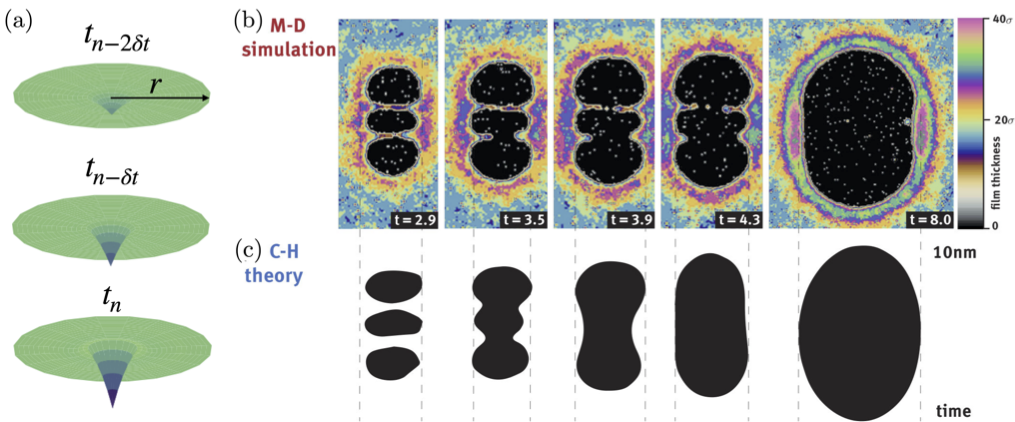}
    \caption{Formation of a nucleus: radial averaged density profile around a rupture site at times closer to its nucleation is revolved around the center to reconstruct the nucleation process, $h_0\sim 11.2$, $r=0$ is the center of the nucleus, and $\delta t =25$ units of MD time. 
    Coalescence of nuclei: simulations; Bottom Panels - Finite Element Analysis using the phenomenological Cahn-Hilliard Theory.  Both are reminiscent of the experimental study of \citet{dolganov2020coalescence}  using free-standing smectic films.}
    \label{fig:MDCH}
\end{figure} 

The time of nucleation (or, rupture formation) inherently possesses a probabilistic nature \citep{vrij1968rupture}, and is subject to fluctuations determined by thermodynamics and other influencing factors. Based on the underlying assumptions, the theoretical prediction of the time to rupture varies from $\tau \approx {h}^3$ (\citet{ruckenstein1974spontaneous}) to $\tau \approx h^5$ (\citet{vrij1968rupture}), where $h$ is the film thickness. 
We discuss these literature in greater details in {\mrr{SI: section 3}}, and compare with the results from current investigation over a range of film thickness. 
Regardless of the time of nucleation (and hence the initial thickness of the film), spontaneous rupture process is preceded by localized thinning, and all rupture sites inevitably undergo this thinning phase ({\mrr{SI: section 3}}). 
Once a film nucleates in our simulations, the highly resolved space-time allow to trace back to the immediate proximity of the nucleation event to capture the formation phase of the hole (i.e., nucleus). The radial averaged local density profile of the center of a nucleus is reconstructed in Fig. \ref{fig:MDCH} (a) where the initiation of local thinning (or dimple formation) is evident ahead of the rupture, i.e., at $t_{n-2\delta t}$, where $\delta =25$ . 
Upon the film's substantial perforation and the subsequent formation of a rim encircling the nucleus, expansion ensues through exponential, and linear growth rates.

While the growth of an isolated nucleus is well understood through the seminal works of Taylor and Culick \citep{Taylor1959, culick1960comments} and subsequent detailed studies \citep{mcentee1969bursting, frankel1969bursting, keller1983breaking, savva2009viscous}, a more commonplace observation is of a film with multiple rupture sites raising a natural question as to how these sites grow and interact. 
The absence of neighbors, or equivalently the consideration of infinite liquid films in previous studies  \citep{savva2009viscous, sunderhauf2002retraction} prohibited the observation of any interaction between propagating rupture sites, and was only realized when the edge effect was accounted for \citep{deka2020revisiting}.
If one considers two neighboring nuclei on a film, these may exhibit three distinct types of behavior in stages: (i) individual uninterrupted growth until when the nuclei come to close proximity of each other, (ii) disturbed growth due to the presence of the neighbor leading to (iii) delayed coalescence. The first stage is predominantly governed by surface tension dynamics. At the second stage, surface tension favors the growth of the nuclei, but coarsening of the bridge between the neighbors retards growth. The third stage requires drainage of the liquid bridge, followed by subsequent thinning and rupture. 

The thickness of this liquid bridge between two neighboring nuclei is notably greater if the nuclei have undergone sufficient growth over time before coming to each other's proximity.
This arises because the liquid molecules from the nuclei accumulate in the surrounding rim, forming a barrier that impedes merging. (Refer to the {\mrr{\textit{SI: section 4}}}  for a scenario where a larger nucleus opts to displace a relatively smaller nucleus rather than coalescing with it.)
However, there are situations where nuclei have expanded extensively and multiplied to the extent that they encounter space limitations, compelling coalescence as the sole energetically favorable course of action to continue to evolve. This phenomenon is illustrated in Figure \ref{fig:MDCH}, where the upper panels showcase results from molecular dynamics simulations depicting the temporal evolution of three artificially generated nuclei, capturing their initial growth, coalescence, and subsequent expansion beyond coalescence. Correspondingly, the lower panels depict analogous outcomes derived from finite element analysis utilizing  Cahn-Hilliard (C-H) theory (see {\mrr{\textit{SI: section 5}}} further discussion). The C-H simulation was initiated by inputting the initial film thickness data with the nuclei (from MD) to affirm the scale invariance of the process. Notably, coalescence commences only when the nuclei encounter constraints impeding further expansion.

\begin{figure}[!th]
    \centering
    \includegraphics[width=0.99\linewidth]{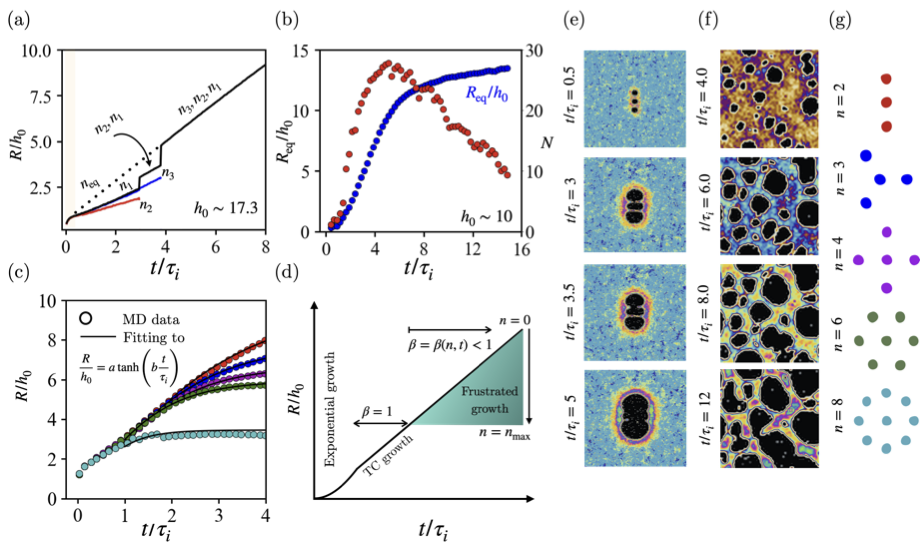}
    \caption{ (a \& e) Growth of three synthetic nuclei ($n_1$, $n_2$, $n_3$) on a film of thickness, $h_0 \sim 17.3$. Snapshots at different instants are shown in panel (e). Initially, each nucleus undergoes exponential growth for a brief period, followed by a phase of linear growth. While all nuclei initially expand at the same rate, the central nucleus experiences hindered growth due to interactions with the adjacent side nuclei, resulting in deceleration (red line). In contrast, the two side nuclei continue their growth at a consistent pace (black: $n_1$, blue: $n_3$). At $t/\tau_i \sim 3$, $n_2$ coalesces with $n_1$, continuing its growth as $n_{c,1,2}$, while $n_3$ maintains separate growth rate until its coalescence at $t/\tau \sim 4$. The newly merged nucleus ($n_{c,1,2,3}$) then resumes unrestrained growth. The dashed line represents the growth trajectory of an equivalent nucleus from the initiation of the linear regime. 
    (b \& f) Growth of the equivalent nucleus (red circles, left y-axis) in a film of thickness $h_0 \sim 10$, where multiple holes spontaneously nucleate - as shown in panel (f). Blue circles in (b )represent the number of nuclei (right y-axis). 
    (c \& g) Effect of the number of neighbors ($n$) on the growth of an expanding central nucleus surrounded by $n$ neighbor nuclei as shown in panel (g), solid lines in (c) denote fitting to the data. (d) Schematic representation of neighbor-effect on the growth of a nucleus. Every case was independently run for at least three times with standard deviation less than $3\%$.
    }
    \label{fig:coals-growth}
\end{figure}
Fig.\,\ref{fig:coals-growth} (a) shows the time-evolution of the radii of three synthetic nuclei (by `synthetic' we mean non-spontaneous nucleus which was induced by applying an exponential force $-$ mimicking the poking of a film in experimental studies \citep{savva2009viscous,bird2010daughter} $-$ in an otherwise stable film, see: methodology) before and beyond coalescence (few representative snapshots are shown in panel e). Initially the nuclei are distant from each other and a very short term exponential growth is noticed followed by the Taylor-Culick linear regime\citep{savva2009viscous, rahman2023non}. However, and as depicted by the red line, the growth of the central nucleus ($n_2$) is affected by the presence of the two expanding side-nuclei ($n_1, n_3$) until it coalesces with one of them. 
Beyond coalescence, the combined nucleus ($n_1$ and $n_2$) maintains linear growth. It is observed from the slopes of the blue (single nucleus, $n_3$) and the black (coalesced $n_1,n_2$, i.e., $n_{c,1,2}$) lines, that the growth rate of the single nucleus, $n_3$ and that of the coalesced nucleus, $n_{c,1,2}$ is the same. 
However, after the final coalescence at around $t/\tau \sim 4$, a slightly steeper slope (of the black line, $n_{c,1,2,3}$) is observed, this is because the growth is no longer restrained by any surrounding nucleus and there is sufficient space for expansion (the film is larger in lateral dimension than in panel e which shows only the area surrounding the nuclei). 
Notably, the dashed line denotes the equivalent radial growth of the three nuclei ($R_\mathrm{eq} = \sqrt{A/\pi}$, with $A = \sum_{i=1}^{3} A_i$) which as well, captures the growth rate of the coalesced nuclei, and thus proves to be a representative measure. More importantly, and in agreement with previous conclusions~\citep{jacobs1998thin,herminghaus1998spinodal}, this implies that the growth of a coalesced nucleus can be extrapolated to the initial nucleation time, provided that growth is hindrance free. 
\citeauthor{jacobs1998thin}\citep{jacobs1998thin} carried out a similar extrapolation in time for an estimation of the nucleation time, but without considering any temporal behavior of the nucleus distribution in spinodal dewetting, and any dynamical instability thereof. 
\citet{herminghaus1998spinodal} also proposed the use of nucleus diameter as a `clock' for the dewetting process. 
However, we will shortly see that such a straightforward extrapolation in time is valid only for a few special cases, and generally imprecise, if not prohibitive.

Figure \ref{fig:coals-growth} (b) provides insight into the growth dynamics of the equivalent nucleus radius, $R_\mathrm{eq}/h_0$ (depicted as red circles corresponding to the left $y$-axis), and the number of nuclei (illustrated as blue circles corresponding to the right $y$-axis) for a relatively thinner film ($h_0 \sim 10$) that ruptures spontaneously; panel (f) presents representative snapshots of this process.
In panel (b), the reference point $t/\tau_i=0$ denotes the time of the first nucleation event. The number of nuclei (blue circles) initially rises during the nucleation phase and subsequently declines from $t/\tau_i \sim 5.8$, indicating the transition to the coalescence regime.
The red circles trace the time evolution of the equivalent nucleus radius, revealing a linear growth during the nucleation regime, followed by a phase of quasi-stagnation where $R_\mathrm{eq}$ experiences marginal increments. 
This contrasts with the observations in panel (a) where the growth rate increased after coalescence. 
The underlying reason for this divergence in growth patterns lies in the available space for expansion. 
For the case in panel (a), the coalesced nuclei have sufficient room to expand, as exemplified by the snapshot at $t/\tau_i=5$ in panel (e). However, for the case in (b), the coalescence of multiple nuclei restricts their individual expansion due to spatial constraints, reminiscent of the middle nucleus in the former case. 
These insights unveil two pivotal aspects of rupture: (i) the growth rate of a nucleus is contingent upon its size and position relative to other nuclei on the film, and (ii) if two nuclei coalesce, the post-coalescence growth rate of the amalgamated nucleus mirrors that of individual nuclei, provided hindrance-free expansion persists. 
This is feasible only when two closely situated nuclei merge during an early stage of the film rupture process, namely before the onset of the coalescence regime. Our observations underscore that a straightforward temporal extrapolation, as discussed earlier, encounters limitations without comprehensive details regarding nucleation dynamics and the spatio-temporal arrangement of the nuclei.

Figure\,\ref{fig:coals-growth} (c) demonstrates how the number of neighbors impact the growth of a central (synthetic) nucleus surrounded by $n$ neighbors. 
The initial radii of all the nuclei, $R (t=0)$, and the initial center-to-center distance between the central and any of the neighbor nuclei were kept constant for all cases considered (the initial conditions are shown in panel g, color of the nuclei in panel g correspond to the symbol color in panel c). 
As $n$ increases, the growth starts deflecting from the linear behavior. One can observe that at $n=6$ (green circles), the growth curve reaches nearly a plateau - meaning the neighbors surround the central  nucleus from all sides inducing sufficient wall effect to suffocate further growth. 
If $n$ is increased beyond 6, the initial growth is shortened and the growth of the central nucleus soon ceases. The solid lines in panel (b) corresponds to fitting of the MD data to an equation of the form, $R/h_0 = a\,\mathrm{tanh \left(bt/\tau_i \right) }$, where $a$ and $b$ are fitting coefficients. 
An expansion of the fitting equation suggests that the frustrated growth can indeed be captured through higher order corrections to the Taylor-Culick law of retraction, i.e., ${R}/{h_0} = c_1 \left({R_\mathrm{TC}}/{h_0}\right) - \mathcal{O}(\chi^n)$, here, $R_\mathrm{TC}$ is the unrestrained radius of the nucleus at time $t$, and $c_1$ is the coefficient of the expansion whose magnitudes is governed by the frustration of growth, and $\mathcal{O}(\chi^n)$ denotes higher order terms in the expansion.
The different stages of nucleus growth including the above mentioned frustrated regime is schematically presented in panel (d). The early-time exponential growth is followed by the Taylor-Culick linear growth, ($U=\beta U_\mathrm{TC}$), with $\beta=1$. Upon sufficient expansion, when a nucleus comes into close proximity of its neighbors, the growth is slowed ($\beta<1$). The extent to which the growth is frustrated depends on the number of neighbors. As $n\to n_\mathrm{max}$ (see {\mrr{\textit{SI: section 4}}} for further details), the frustration is maximized and growth is ceased.

\begin{figure}[!t]
\centering 
\includegraphics[width=0.85\linewidth]{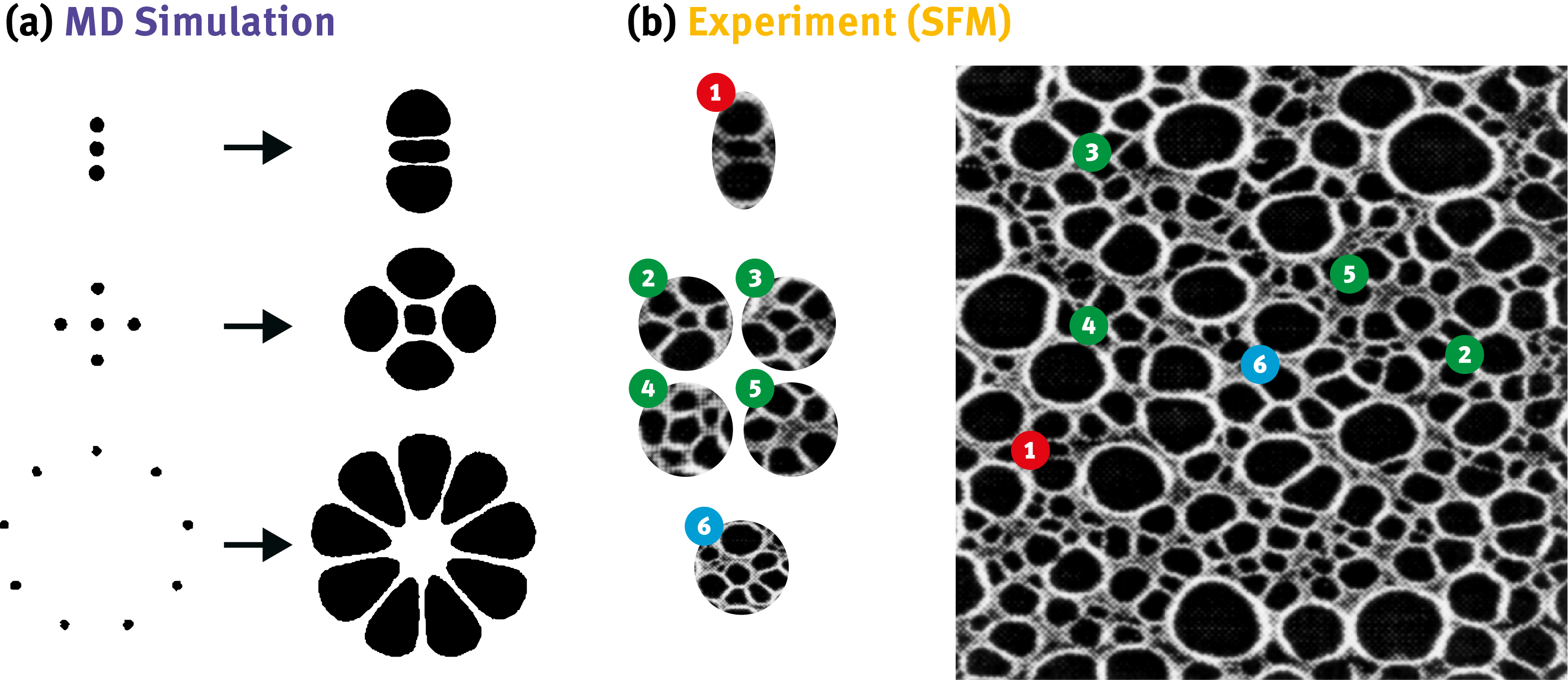}
\caption{ (a) Patterns observed in current investigations through varied synthetic perturbation. The left images depict the initial conditions, while the right images capture snapshots during the later stages of rupture.  (b) The patterns observed in panel a, are compared to the those observed in the scanning force microscopy (SFM) image (adapted from panel c of Figure 1 in \citeauthor{thiele1998dewetting}, \citeyear{thiele1998dewetting}, with permission from the authors and the publishers) of a collagen film on a solid substrate. Similar looking patterns in MD and experiments are marked by identical numbers.}
\label{fig:patterns}
\end{figure}

As time unfolds, the resulting grown-up patterns show striking similarity with patterns observed in earlier experimental studies.  
Figure \,\ref{fig:patterns} compares some of the patterns observed in the present investigation with the experimental findings by \citet{thiele1998dewetting}. 
The left column on panel (a) presents the initial stages of synthetic nuclei, which are allowed to evolve over time, the corresponding later stages are displayed on the right column of the panel.
Panel (b) shows similar looking patterns found through scanning force microscopy (SFM)  of collagen film rupturing on a solid substrate, revealing a diverse array of rupture sites with varying shapes and sizes. 
This congruence between our simulations and experimental observations underscores the effectiveness of molecular dynamics (MD) simulations as a robust tool for dissecting the intricate dynamics of liquid film rupture.

\section*{Conclusions}
To summarize, the present investigation unravels the complete life cycle of a black film, delving into its inception, evolution, and ultimate fragmentation with intricate insights into these phases. Our findings contrast the conventional understanding of dewetting or rupture in pure films, revealing that the unfolding of film morphologies - commonly characterized as spinodal and heterogeneous nucleation - is fundamentally identical except for their respective timescales. The highly resolved time in our investigation also discloses a ‘film-pinching’ phase preceding the exponential growth of rupture. A brief but discernible window of determinism, which emerges amidst the stochastic nature of rupture dynamics, indicates the liquid’s inherent structure as the gateway to predicting rupture. Following the evolution of the nucleated sites through exponential and linear growths, the late stage is marked by considerable frustration in growth, resulting in delayed coalescence and coarsening. In essence, our study sheds light on the intricacies of black-film rupture dynamics, providing valuable insights into the underlying mechanisms governing this phenomenon. These not only expand our fundamental understanding of thin film behavior but also open doors to potential applications in various scientific and technological domains.

\pagebreak
\section*{Methodology}

\noindent \textbf{Molecular dynamics simulations:}
The model films studied here are composed of single component
Lennard-Jones (LJ) particles, i.e., $U_{LJ} = 4\epsilon [(\sigma/r)^{12} - (\sigma/r)^6]$, which are free from electrostatic interactions. The only force that can have a destabilizing effect arises from the van der Waals attraction term in the potential, $U_{LJ}$, which competes with the surface tension. 
Simulations were carried out employing the extensively validated and verified Flowmol MD code~\citep{smith2013coupling}. 

The initial simulation domain was represented by a cubic box with dimensions $L_x = 76.19$ and $L_y = L_z = 609.56$ in LJ units, translating to film dimensions approximately 207 nm wide and with a depth ($h_0$) of less than 5 nm. Throughout this paper, all measurements and quantities are expressed in LJ units.
The central $20\%$ of the simulation box along the $x$ axis was assigned as the liquid phase, with a targeted density of $\rho\approx 0.7$, while the remaining region was designated as the vapor phase with a lower density of $\rho\approx 0.01$. 
A cohesive central liquid film, consisting of around $\sim 0.4 - 1.8$ million identical atoms, coexists harmoniously with the surrounding vapor phase at an equilibrium state. 
The equilibration process of the film was carefully conducted under a controlled temperature of $T=0.78\pm0.03$. 
To explore films with different thicknesses, the dimension in the $x$ direction was systematically varied.

The absence of surfactants or any impurity allowed us to discard the possibility of having any external species and their uncharted interactions. Apart from the fundamental appeal, investigations of such pure films is directly relevant in gaining a deeper understanding of  metallic foams which are always surfactant-free \citep{anderson2010foam}. \newline

\noindent {\textit{Synthetic nucleation:}}
To initiate synthetic nucleation, we applied an external force, as in Eq.\,(\ref{eq:force_ext}), to mimic the popular experimental film-puncturing procedure~\citep{savva2009viscous,bird2010daughter,oratis2020new}. 
\begin{align}
        F_\mathrm{ext.} = \left\{
            \begin{array}{ll}
                 F \left( 1 + \mathrm{e}^{-1/r^2} \right) & \quad r \leq R_0 \\
                0 & \quad \mathrm{otherwise},
            \end{array}
        \right.
        \label{eq:force_ext} 
\end{align}
where, $R_0$ is the initial radius of the hole. 
The forcing time was only a few hundred time-steps to confirm that a stable hole is created, and $R_0$ was kept sufficiently small of the order of the film thickness.
Prior investigations have demonstrated that this artificial nucleation process does not impact the growth dynamics of the nuclei~\citep{rahman2023non}.

Following the equilibration (and synthetic nucleation), the system was further simulated using the micro-canonical (NVE) ensemble. \\

\noindent \textbf{Finite element analysis employing Cahn-Hilliard Theory:} \newline 
The finite element analysis for the Cahn-Hilliard theory uses a mixed formulation which recast the fourth-order equation into two coupled second-order equations. This bypasses the $C^1$-continuous requirement of the Galerkin method and has shown \cite{Zhang2013,Kaessmair2016} to give comparable accuracy to the $C^1$-continuous methods whilst being less computationally expensive. To realise the finite element formulation, the solution of a system
state is given via a interpolation function $A(\mathbf{x})=\sum_{i=1}^{N}A_{i}N_{i}(\mathbf{x}),$
where $N_{i}$ is the finite-element basis, and $A_{i}$ is the coefficient
for each triangular elements with the indices $i=1,\ldots,N$. A standard rectangular mesh is used whereby the initial conditions are imported from a state of the molecular dynamics simulations. The FEM package FENiCs \cite{Alnaes2015} is used for the calculation  and the time discretisation uses the Crank-Nicolson method. The Newton-Krylov solvers based on PETSc's SNES module is used with the discretisations in space and time solved using the general minimal residual method (GMRES). Each iteration is solved to a relative tolerance of $10^{-6}$. The solution process scales well with multiple cores using the MPI routine. 

\pagebreak 

\bibliographystyle{unsrtnat}
\bibliography{ref_spont}


\section*{Acknowledgments}
M.R.\,R. thanks Shell and the Beit Trust for PhD funding through a Beit Fellowship for Scientific Research. L.S. thanks the Engineering and Physical Sciences Research Council (EPSRC) for a Postdoctoral Fellowship (EP/V005073/1). J.P.E. was supported by the Royal Academy of Engineering (RAEng) through their Research Fellowships scheme. D.D. acknowledges a Shell/RAEng Research Chair in Complex Engineering Interfaces and the EPSRC for an Established Career Fellowship (EP/N025954/1). \newline

\section*{Author Contributions}
D.D. and L.S. acquired the funding; D.D., E.R.S., D.M.H., L.S. and J.P.E. conceived the problem and supervised the research; M.R.R. performed MD simulations; L.S. performed FE simulations; M.R.R. analyzed the results; M.R.R. and L.S. wrote the first draft with inputs from E.R.S.; all authors discussed the results and edited the manuscript. 

\section*{Competing Interests} 
The authors declare no competing interests.

\section*{Correspondence}
Correspondence to: Muhammad Rizwanur Rahman 


\section*{Data Availability}
Data supporting the findings of this manuscript are available upon reasonable request to the corresponding author.

\section*{Additional information}
Supplementary information available online.

\end{document}